\documentclass[osajnl,twocolumn,showpacs,superscriptaddress,10pt]{revtex4-1} %% use 11pt for Applied Optics
\usepackage{amsmath,amssymb,graphicx}
\usepackage{revsymb4-1}
\usepackage{bm}
\usepackage{braket}

\begin{document}

\title{Dynamics of entangled states in squeezed reservoirs}
%\subtitle{Dynamics of entangled states in squeezed reservoirs}
\author{Khalid Aloufi}
\author{Smail Bougouffa}
\email{sbougouffa@hotmail.com and sbougouffa@taibahu.edu.sa}
\affiliation{Department of Physics, Faculty of Science, Taibah University, P.O. Box 30002, Madinah~41481, Saudi Arabia}
\author{Zbigniew Ficek}
\email{zficek@kacst.edu.sa}
\affiliation{National Center for Mathematics and Physics, KACST, P.O. Box 6086, Riyadh 11442, Saudi Arabia}

\date{\today}

\begin{abstract}
Dynamics of entangled states of two independent single-mode cavities in squeezed reservoirs is investigated in the context of matching of the correlations contained in the entangled states to those contained in the squeezed reservoir. We illustrate our considerations by examining the time evolution of entanglement of single and double excitation NOON and EPR states. A comparison is made when each cavity is coupled to own reservoir or both cavities are coupled to a common reservoir. It is shown that the evolution of the initial entanglement and transfer of entanglement from the squeezed reservoir to the cavity modes depend crucially on the matching of the initial correlations to that contained in the squeezed reservoir. In particular, it is found that initially entangled modes with correlations different from the reservoir correlations prevent the transfer of the correlations from the squeezed field to the modes. In addition, we find that the transient entanglement exhibits several features unique to quantum nature of squeezing. In particular, we show that in the case of separate squeezed reservoirs a variation of the decay time of the initial entanglement with the squeezing phase is unique to quantum squeezing. In the case of a common reservoir a recurrence of entanglement occurs and we find that this feature also results from the reservoir correlations unique to quantum squeezing. There is no revival of entanglement when the modes interact with a classically squeezed field.
\end{abstract}


\maketitle

\section{Introduction}

Interest in nonclassical (quantum) aspects of squeezed light has been greatly stimulated by the need to generate entangled pairs of photons~\cite{nil,df04,FT02,SLF10}. The experimental generation of squeezed light in a variety of phase-dependent nonlinear optical processes has provided the opportunity to demonstrate fundamental quantum correlations typical for entangled systems~\cite{SHYMV85,SLPDW86,GPEKP95}. The quantum correlations in squeezed light are manifested by the emission of correlated pairs of photons. Entanglement is a consequence of the quantum correlations between two systems and similar to squeezing is a nonclassical phenomenon which have played an important role in developing our understanding of the quantum world. It is an essential resource for various quantum algorithms such as quantum teleportation~\cite{BBCJPW93}, quantum dense coding~\cite{BW92}, quantum cryptography~\cite{E91} and quantum computing~\cite{BDEJ95}.

Important for practical applications of entanglement is the problem of transferring entangled states between distant systems, in particular, the manner the entanglement evolves in time. It is well known that entangled states are fragile to decoherence which has a destructive effect on entanglement and may disentangle an initially entangled system in a finite time~\cite{YE04,YE06,YE106,YYE06,YoE08,EY09,AMHSWRSD07,FT06,FT08}. For the simplest systems involving two qubits, the evolution of an initial entangled state depends strongly on the statistics of the reservoir field surrounding the qubits. For example, it has been shown that for non-zero temperature reservoir, the disentanglement of an entangled system in a finite time, known as the entanglement sudden death, is the standard feature of the evolution, and only in the limit of zero temperature the asymptotic decay of entanglement is possible~\cite{TIBZ10,AILZ07,TIAZ08,B10,BH11,B11}.

Dynamics of entangled states in a correlated reservoir such as a squeezed vacuum may be significantly different then that in a thermal reservoir due to the presence of quantum correlations~\cite{MWF12,TF04,MO07,KL10,Y07}. In this paper we discuss the problem in the context of matching of the correlations contained in the entangled states to those contained in the squeezed reservoir. We compare the time evolution of entanglement of initial single and double excitation NOON and EPR states. In addition, we explore purely nonclassical features in the entanglement evolution and transfer in the sense that they arise from the quantum nature of the squeezed reservoir. In other words, we require that for a squeezed reservoir with classically correlated modes these features cease to exist. Nonclassical effects in the radiative properties of atoms decaying into a squeezed reservoir have been observed in a number of experiments~\cite{GP95,PG97,TG98,MW13}.

The paper is organized as follows. In section~\ref{sec2} we introduce the model and formulate the master equation for the density operator of two cavity modes each interacting with a squeezed reservoir or both interacting with a common reservoir. In section~\ref{sec3} we define entanglement measures and the basis states for singly and doubly excited states of the system. We use concurrence to quantify entanglement between the modes for the case of singly excited states and logarithmic negativity to quantify entanglement of doubly excited states. In section~\ref{sec4} we explore the dynamics of the initial NOON and EPR states
in separate and common squeezed vacuum reservoirs. We pay particular attention to features that are unique for quantum nature of the squeezed field. In addition, we illustrate the role of the relative phase between the initial state of the system and the squeezed reservoir in the evolution of entanglement. We find that the evolution of NOON states is not sensitive to the squeezing phase. In contrast, EPR states exhibit a strong dependence on the phase which in the case of separate reservoirs modifies the decay time of the initial entanglement. In the case of a common reservoir, the phase influences the time at which the transfer of entanglement from the squeezed reservoir to the modes starts to begin. Finally, in section~\ref{sec5} we conclude with a short summary of our results.

\section{Model and description}\label{sec2}

\begin{figure}[h]
\center{\includegraphics[width=1.\columnwidth]{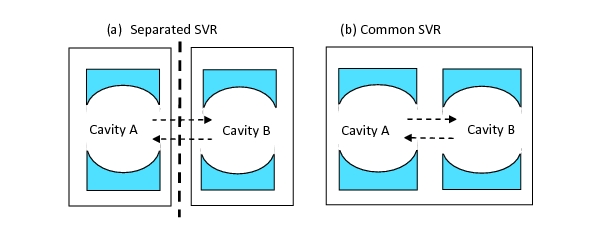}}
\caption{Two independent systems of identical single-mode cavities containing initial entangled fields. (a) The cavity modes do not have
directional interaction with each other but independently interact with their local squeezed vacuum reservoir. (b) Both cavities interact with the same squeezed vacuum reservoir.}
\label{Fig1}
\end{figure}

We consider a system composed of two single-mode cavities interacting with a broadband squeezed vacuum field. The squeezed field whose bandwidth is much larger than the cavity decay rate $\kappa$ is injected into cavities through one of their mirrors and serves as a reservoir to the cavity mode. We study separately two cases, shown in Fig.~\ref{Fig1}. In the first case, we assume that each cavity is coupled to its own squeezed vacuum reservoir and the reservoirs are independent of each other. This kind of squeezed field is obtained from a degenerate parametric process and is refer to as single-mode squeezing. In the second case, both cavities are assumed to be coupled to a single (common) squeezed vacuum reservoir. This kind of squeezed field is obtained from a non-degenerate parametric process and is refer to as two-mode squeezing.

In the case when each cavity interacts with its own reservoir, the interaction of the cavity modes with the multimode reservoir field is described by the Hamiltonian, which in the interaction picture and under the rotating-wave approximation, can be written as
\begin{eqnarray}
  H = \hbar\sum_{j=A,B}\sum_{\bf{k}}\left[ g_{\bf{k}} b_{\bf{k}}^{\dag}a_{j}
  e^{-i(\nu-\nu_{{k}})t}+ g^{*}_{\bf{k}}a_{j}^{\dag}b_{\bf{k}}e^{i(\nu-\nu_{{k}})t}\right] ,\label{e1}
    \nonumber\\
\end{eqnarray}
where $a_{j}$ and $a^{\dag}_{j}\, (j=A,B)$ are the bosonic annihilation and creation operators of the mode of the $j$th cavity, $b_{\mathbf{k}}$ and $b_{\mathbf{k}}^{\dag}$ are the annihilation and creation operators of mode~$k$ of frequency $\nu_{k}$ of the reservoir, and $g_{\mathbf{k}}$ is the coupling constant of the cavity modes to the $k$ mode of its reservoir. For simplicity, we assume that the frequency $\nu$ of the cavity modes and the coupling constants $g_{\mathbf{k}}$ are the same for each cavity.

A usual way of illustrating the effect of a reservoir on the dynamics of a given system is to consider the master equation for the density operator $\rho$ of the system. When the modes are coupled to independent squeezed vacuum fields the master equation for the density operator $\rho$, under the standard Born-Markov approximations, is of the form~\cite{SZ97}
\begin{eqnarray}
 \frac{d\rho}{dt} &=&\frac{1}{2}\kappa \sum_{j=A,B}\Bigg[(N_{j}+1)\left(2a_{j}\rho a_{j}^{\dag} -a_{j}^{\dag}a_{j}\rho -\rho a_{j}^{\dag}a_{j}\right)\nonumber\\
 &&+ N_{j}\left(2a_{j}^{\dag}\rho a_{j}-a_{j}a_{j}^{\dag}\rho -
   \rho a_{j}a_{j}^{\dag} \right)\nonumber\\
  &&- M_{j}\Big(2a_{j}\rho a_{j}-a_{j}a_{j}\rho -\rho a_{j}a_{j} \Big)\nonumber\\
   &&- M_{j}^{*}\left(2a_{j}^{\dag}\rho a_{j}^{\dag}-a_{j}^{\dag}a_{j}^{\dag}\rho -
   \rho a_{j}^{\dag}a_{j}^{\dag}\right)\Bigg] ,\label{e2}
\end{eqnarray}
where $\kappa$ is the decay rate of the modes, assumed the same for both cavities, and $N_{j}$ is the number of photons in the~$j$th mode.
The first term in Eq.~(\ref{e2}) represents decay of the $j$th mode with the rate $\kappa(N_{j}+1)$, the second term represents an incoherent pumping of the modes with rate~$\kappa N_{j}$. The third and fourth terms represent correlations between photons with the degree $M_{j}=|M_{j}|\exp(-i\theta_{j})$. The parameter $|M_{j}|$ determines the degree of two-photon correlations inside the $j$th mode. Therefore, it is referred to as the degree of a single-mode squeezing.

It should be pointed out that there is a clear distinction between classical and nonclassical (quantum) regimes for squeezing~\cite{df04} that involve classical and quantum correlations, respectively. The regimes are determined by the degree of correlations $|M|$. The classical regime for squeezing, often called to as a classical squeezing, is determined by $0<|M_{j}|\leq N_j$, whereas $N_{j}<|M_{j}|\leq \sqrt{N_j(N_j+1)}$ indicates the nonclassical regime for squeezing, often called to as quantum squeezing. Note that the field with $M=0$ and $N>0$ is a thermal field while the field with $M=0, N=0$ is the ordinary vacuum field. The parameters $M$ and $N$ can also be expressed in terms of the squeezing parameter $r$, $M=\sinh r \cosh r$ and $N=\sinh^{2}r$.

In the case when the cavities are coupled to a common reservoir, the interaction Hamiltonian between the cavity and the reservoir modes is given by
\begin{eqnarray}\label{3}
  H = \hbar\sum_{\bf{k}}&\Big[g_{\bf{k}} b_{\bf{k}}^{\dag}(a_{A}+a_{B})
  e^{-i(\nu-\nu_{{k}})t} \nonumber \\
  &+ g_{\bf{k}}^{*}(a_{A}^{\dag}+a_{B}^{\dag})b_{\bf{k}}
    e^{i(\nu-\nu_{{k}})t}\Big] .
\end{eqnarray}
For this case, the master equation for the density operator~$\rho$ of the system takes the form~\cite{SZ97}
\begin{eqnarray}\label{4}
 \frac{d\rho}{dt} &=& \frac{1}{2}\kappa \sum_{i=A,B}\Bigg[(N_{i}+1)\left(2a_{i}\rho a_{i}^{\dag} -a_{i}^{\dag}a_{i}\rho -\rho a_{i}^{\dag}a_{i}\right)\nonumber\\
 &&+ N_{i}\left(2a_{i}^{\dag}\rho a_{i}-a_{i}a_{i}^{\dag}\rho -
   \rho a_{i}a_{i}^{\dag} \right)\Bigg]\nonumber\\
   &-&\frac{1}{2}\kappa\sum_{i\neq j=A,B}\Bigg[ M_{ij}\left(2a_{j}\rho a_{i} -a_{i}a_{j}\rho -\rho a_{i}a_{j}\right)\nonumber\\
   &+& M_{ij}^{\ast}\left(2a_{j}^{\dag}\rho a_{i}^{\dag} -a_{i}^{\dag}a_{j}^{\dag}\rho -\rho a_{i}^{\dag}a_{j}^{\dag}\right)\Bigg] ,
\end{eqnarray}
where $M_{ij}=|M_{ij}|\exp(-i\theta)$ with $|M_{ij}|=\sqrt{N_{i}(N_{j}+1)}$. The parameter $M_{ij}$ determines the degree of two-photon correlations between the modes and, therefore, is referred to as the degree of two-mode squeezing. As before for the separate reservoirs, $M_{ij}<N_{i}$ corresponds to a classically squeezed field while  $M_{ij}>N_{i}$ corresponds to a quantum squeezed field.

We are interested in the evolution of an entangled state in a squeezed vacuum field. Therefore, we assume that in the time period before $t=0$ the cavity modes are prepared in a superposition state and then at $t=0$ we allow the modes to interact with the squeezed vacuum field. In practice, the preparation of the cavity modes in an entangled state can be done by using a parametric downconversion source and a beamsplitter. It is well know that the output of a parametric downcoverter is composed of strongly correlated pairs of photons that can be used for a simultaneous excitation of the input modes of the beamsplitter~\cite{HM85,HO87}. The output of the downconverter can also be used as a source of single photons~\cite{LO05}. Alternatively, one can use quantum dots or nitrogen vacancy centres in diamond as sources of single photons~\cite{MI00,SH07,MM12,CJ14}. Thus, by exciting one of the input modes of the beamsplitter with the output of a single photon source, an $n=1$ NOON state can be created between the output modes of the beamspitter. Similarly, by exciting both input modes of the beamsplitter simultaneously with the output modes of the downconverter, an $n=2$ NOON state can be created between the output modes of the beamsplitter.

We consider the evolution of two different sets of initial entangled states at $t=0$. In the first, the modes are assumed to be prepared in a NOON entangled state
\begin{eqnarray}
\ket{\Psi_{AB}(0)} = \cos\alpha\ket{0_{A}}\ket{n_{B}} + e^{-i\psi} \sin\alpha\ket{n_{A}}\ket{0_{B}} .\label{e5}
\end{eqnarray}
In the second, the modes are assumed to be prepared in an EPR entangled state
\begin{eqnarray}
\ket{\Psi_{AB}(0)} = \cos\alpha\ket{0_{A}}\ket{0_{B}}+e^{-i\psi}\sin\alpha\ket{n_{A}}\ket{n_{B}} ,\label{e6}
\end{eqnarray}
where $n_{j}\, (j=A,B)$ is the number of excitations (photons) present in the mode of the $j$th cavity. Note that the initial states~(\ref{e5}) and~(\ref{e6}) are defined for an arbitrary phase $\psi$, which may differ from the phase $\theta$ of the squeezed field. Thus, one can monitore the variatiation of entanglement with the phase by fixing $\psi$ and evaluating the concurrence as a function of $\theta$.

In what follows, we limit the number of excitations present in the cavity modes to the cases of $n_{j}=1$ and $n_{j}=2$. This is justified by assumining that the average number of photons in the squeezed vacuum is small, $N_{i}\ll 1$. In order show it more explicitly, we write the master equation in the photon-number representation from which we find that the steady state for the diagonal elements $P_{n}=\rho_{n,n}$ obeys the recurrence relation
\begin{eqnarray}
nNP_{n-1} &-& (2nN+N+n)P_{n} \nonumber\\
&+& (N+1)(n+1)P_{n+1} +MC_{n} =0, \label{e7}
\end{eqnarray}
where
\begin{eqnarray}
C_{n} &=& \frac{1}{2}\left[ \sqrt{(n+1)(n+2)}(\rho_{n+2,n} +\rho_{n,n+2})\right. \nonumber\\
&-&\left. 2\sqrt{n(n+1)}\left(\rho_{n+1,n-1}+\rho_{n-1,n+1}\right)\right. \nonumber\\
&+&\left. \sqrt{n(n-1)}(\rho_{n-2,n} -\rho_{n,n-2})\right] ,
\end{eqnarray}
with $N=N_{1}=N_{2}$.

From the recurrence relation, one can easily show that in the case of a weak squeezed field with $N=0.1$ and $M=\sqrt{N(N+1)}$, the population distribution is essentially the same as that for a thermal field. In particular, the ratio $R_{n}=P_{n}/P_{0}$ of the populations of the $n$th state to that of the $n=0$ state is found to vary with $n$ as: for $n=1$, $R_{1}=0.083$,  for $n=2$, $R_{2}=0.008$ and for $n=3$, $R_{3}=0.002$.
Clearly, the population of the states $n\geq 3$ is very small and can be neglected. This shows that essential for the dynamics are states with the number of excitations $n=0, n=1$ and $n=2$.

\section{Entanglement measures}\label{sec3}

Our objective is to study the effect of the correlations present in the squeezed reservoir on the dynamics of entangled states. In order to study this effect we need entanglement measures. For this purpose we adopt two commonly used measures, the concurrence and logarithmic negativity. We adopt the concurrence to calculate the evolution of entanglement of single excitation states, whereas the logarithmic negativity is used to evaluate entanglement of doubly exited state. The reason of using two different measures is in different dimensions of the $n=1$ and $n=2$ systems.

\subsection{Singly excited states}

In the case in which there can be maximally a single excitation present in each mode the Hilbert space of the system can be spanned in terms of four product states
\begin{eqnarray}
\ket{1} &= \ket{0_{A}}\ket{0_{B}} ,\quad \ket{2} =\ket{0_{A}}\ket{1_{B}} , \nonumber\\
\ket{3} &= \ket{1_{A}}\ket{0_{B}} ,\quad \ket{4} =\ket{1_{A}}\ket{1_{B}} .\label{e4a}
\end{eqnarray}
We see that the space of the system corresponds to that of a $2\times 2$ system.
In practice, the single excitation states of the cavity modes can be generated by sending an excited two-level atom successively through the cavities, which initially were in the vacuum state $\ket{0_{A}}\ket{0_{B}}$.

In the basis of the product states (\ref{e4a}) the density matrix~$\rho$, for both cases of separate and common reservoirs, takes the form
\begin{equation}
    \rho =\left(%
\begin{array}{cccc}
  \rho_{11} & 0 & 0 & \rho_{14} \\
  0 & \rho_{22} & \rho_{23} & 0 \\
  0 & \rho_{32} & \rho_{33} & 0 \\
  \rho_{41} & 0 & 0 & \rho_{44} \\
\end{array}
\right) .\label{15}
\end{equation}

Since the Hilbert space of the system corresponds to that of a $2\times 2$ system, we can use the concurrence to evaluate entanglement between the modes. The concurrence is defined by~\cite{W98,W06}
\begin{equation}\label{11}
    C(t) = {\rm max}\left(0,\sqrt{\lambda_{1}}-\sqrt{\lambda_{2}}-\sqrt{\lambda_{3}}-\sqrt{\lambda_{4}}\right) ,
\end{equation}
where $\lambda_{i}$ are the eigenvalues of the non-Hermitian matrix~$\rho\tilde{\rho}$ arranged in decreasing order of their magnitudes, and
\begin{equation}\label{12}
    \tilde{\rho} = (\sigma_{y}^{A} \otimes\sigma_{y}^{B} )\rho^{*} (\sigma_{y}^{A}\otimes\sigma_{y}^{B}) ,
\end{equation}
in which $\sigma_{y}$ is the Pauli matrix. The concurrence varies between $C=0$ for a separable state and
$C=1$ for a maximally entangled state, and $0<C<1$ for mixed quantum states.

Given the simple form of the density matrix (\ref{15}) we can calculate the concurrence analytically, and find a quite simple expression
\begin{equation}
    C(t) = {\rm max}\left(0,\tilde{C}_{1}(t),\tilde{C}_{2}(t)\right) ,\label{16}
\end{equation}
where
\begin{eqnarray}
  \tilde{C}_{1}(t) &=& 2\left(|\rho_{23}(t)|-\sqrt{\rho_{11}(t)\rho_{44}(t)}\right) ,\label{17} \\
   \tilde{C}_{2}(t) &=& 2\left(|\rho_{14}(t)|-\sqrt{\rho_{22}(t)\rho_{33}(t)}\right) .\label{18}
\end{eqnarray}
From Eqs.~(\ref{17}) and (\ref{18}) it is clear that the necessary condition for entanglement is that either $|\rho_{23}(t)|$ or $|\rho_{14}(t)|$ is different from zero. Note that $\rho_{23}$ accounts for entanglement of the NOON state, Eq.~(\ref{e5}), whereas $\rho_{14}$ accounts for entanglement of the EPR state, Eq.~(\ref{e6}).

\subsection{Doubly excited states}

When up to two quanta of excitation could be present in each mode, the Hilbert space of the system can be   spanned in terms of nine product state vectors
\begin{eqnarray}\label{9}
& \ket{1}= \ket{0_{A}}\!\ket{0_{B}},\ \ket{2}= \ket{0_{A}}\!\ket{1_{B}},\ \ket{3}=\ket{0_{A}}\!\ket{2_{B}} , \nonumber\\
 & \ket{4}= \ket{1_{A}}\!\ket{0_{B}},\ \ket{5}= \ket{1_{A}}\!\ket{1_{B}},\ \ket{6}= \ket{1_{A}}\!\ket{2_{B}} , \nonumber\\
 & \ket{7}= \ket{2_{A}}\!\ket{0_{B}},\ \ket{8}= \ket{2_{A}}\!\ket{1_{B}},\ \ket{9}= \ket{2_{A}}\!\ket{2_{B}} .\label{e14}
\end{eqnarray}
In this case, the Hilbert space corresponds to that of a $3\times 3$ system. For this reason, we cannot use the concurrence to evaluate entanglement. Instead, we take as adequate the logarithmic negativity, defined as~\cite{VW02,LCK03,P05}
\begin{equation}
    \mathcal{N}=\log_{2}\|\rho^{T_{B}}\|_{1} ,\label{19}
\end{equation}
where $\rho^{T_{B}}$ is the partial transpose of the density matrix and $\|.\|_{1}$ is the trace norm
\begin{equation}\label{20}
    \|\rho^{T_{B}}\|_{1}=1+2|\sum_{l}\mu_{l}| ,
\end{equation}
in which $\mu_{l}$ are the negative eigenvalues of $\rho^{T_{B}}$. The negativity varies between $\mathcal{N}=0$ for separable states and $\mathcal{N}=1$ for maximally entangled state.

If we arrange the states (\ref{e14}) in the following order
\begin{equation}
 \Big \{\ket{1},\ket{3},\ket{7},\ket{9},\ket{2},\ket{8},\ket{4},\ket{6},\ket{5}\Big\}
\end{equation}
we find that in the case of separate reservoirs, the matrix representation of the density operator takes a block diagonal form
\begin{equation}
    \rho =\left(%
\begin{array}{ccccccccc}
  \rho_{11} & \rho_{13} & \rho_{17}& \rho_{19} & 0 & 0 & & 0 & 0\\
 \rho_{31} & \rho_{33} & \rho_{37} & \rho_{39} & 0 & 0 & 0 & 0 & 0\\
 \rho_{71} & \rho_{73}& \rho_{77} & \rho_{79} & 0 & 0 &0 & 0 & 0\\
 \rho_{91} & \rho_{93} & \rho_{97} & \rho_{99} & 0 &0 & 0 & 0 & 0\\
  0 & 0 &0 & 0 & \rho_{22} & \rho_{28}& 0 & 0 & 0\\
 0 & 0 & 0 & 0 & \rho_{82} & \rho_{88} & 0 & 0 & 0\\
  0 & 0 & 0 & 0 & 0 & 0 & \rho_{44} & \rho_{46} & 0\\
 0 & 0 & 0 & 0 & 0 & 0 & \rho_{64} & \rho_{66} & 0\\
  0& 0 &0 & 0 & 0 & 0 & 0 & 0 & \rho_{55}\\
\end{array}%
\right) .
\end{equation}
It is easily verified that entanglement between the modes can come only from the coherences involved in the
$4\times 4$ matrix since they correlate states of different modes. The coherences involved in the remaining $2\times 2$ blocks cannot produce entanglement because they correlate states of the same mode.

For the case of the common reservoir we arrange the product states in the following order
\begin{equation}\label{38}
    \Big\{\ket{1},\ket{5},\ket{9},\ket{2},\ket{6},\ket{4},\ket{8},\ket{3},\ket{7}\Big\} ,
\end{equation}
and find that the matrix representation of the density operator takes a block diagonal form
\begin{equation}\label{39}
    \rho =\left(%
\begin{array}{ccccccccc}
  \rho_{11} & \rho_{15} & \rho_{19} & 0 & 0 & 0 & 0 & 0 & 0\\
 \rho_{51} & \rho_{55} & \rho_{59} & 0 & 0 & 0 & 0 & 0 & 0\\
  \rho_{91} &\rho_{95} & \rho_{99} & 0 & 0 & 0 & 0 & 0 & 0\\
 0 & 0 & 0 & \rho_{22} & \rho_{26} &0 & 0 & 0 & 0\\
  0 & 0 &0 & \rho_{62} & \rho_{66} & 0 & 0 & 0 & 0\\
 0 & 0 & 0 & 0 & 0 & \rho_{44} & \rho_{48} & 0 & 0\\
  0 & 0 & 0 & 0 & 0 & \rho_{84} & \rho_{88} & 0 & 0\\
 0 & 0 & 0 & 0 & 0 & 0 & 0 & \rho_{33} & \rho_{37}\\
  0 & 0 & 0 & 0 & 0 & 0 & 0 & \rho_{73} & \rho_{77}\\
\end{array}%
\right) .
\end{equation}
By inspection of the matrix (\ref{39}) one can easy find that the coherences involved in each block correlate states of different modes, so that they can produce entanglement.

\section{Dynamics of the entangled states}\label{sec4}

We now proceed to illustrate the results for the dynamics of entanglement of the initial NOON and EPR states of the cavity modes. We consider separately the cases of individual and common squeezed vacuum reservoirs to which the modes are coupled.

\subsection{Dynamics in the ordinary vacuum reservoir}

Let us first briefly discuss the evolution of the entangled states in ordinary vacuum reservoirs, $M=N=0$. In a such limit, it is sufficient either to work with the case of separate reservoirs or with the case of the common reservoir. The results will serve as a reference for that obtained for the evolution in squeezed reservoirs.

For the modes initially prepared in the $n=1$ NOON state, nonzero density matrix elements at $t=0$ are, $\rho_{22}(0) = \cos^{2}\alpha, \rho_{33}(0) = \sin^{2}\alpha, \rho_{23}(0)=\rho_{32}^{\ast}(0) = e^{i\psi}\sin\alpha\cos\alpha$. Using the master equation (\ref{e2}) we easily find that the density matrix elements undergo an exponential decay
\begin{eqnarray}
\rho_{11}(t) &=& 1-e^{-\kappa t},\ \rho_{22}(t)=e^{-\kappa t}\cos^{2}\alpha ,\nonumber\\
\rho_{23}(t) &=& \rho_{32}^{\ast}(t)= e^{-\kappa t}e^{i\psi}\cos\alpha\sin\alpha ,\nonumber\\
\rho_{33}(t) &=& e^{-\kappa t}\sin^{2}\alpha ,\label{e21}
\end{eqnarray}
which leads to a simple time evolution of the concurrence
\begin{equation}
     C(t) = {\rm max}\{0,\tilde{C}_{1}(t)\} = {\rm max}\{0,|\sin2\alpha|e^{-\kappa t}\} .\label{e22}
\end{equation}
Clearly, $\tilde{C}_{1}(t)$ is positive for all values of $\alpha$ and exhibits a simple exponential decay in time. Thus, the initially entangled modes remain entangled over the entire decay time.

In the case of an initial $n=1$ EPR state, nonzero density matrix elements at $t=0$ are, $\rho_{11}(0) =\cos^{2}\alpha, \rho_{44}(0)=\sin^{2}\alpha, \rho_{14}(0)=\rho_{41}^{\ast}(0)=e^{i\psi}\sin\alpha\cos\alpha$. The resulting time evolved density matrix elements are then given by
\begin{eqnarray}
\rho_{11}(t) &=& 1-\left(2 - e^{-\kappa t}\right)e^{-\kappa t}\sin^{2}\alpha , \nonumber \\
\rho_{22}(t) &=& \rho_{33}(t) = (1-e^{-\kappa t})e^{-\kappa t}\sin^{2}\alpha , \nonumber\\
\rho_{14}(t) &=& \rho_{41}^{\ast}(t) = e^{-\kappa t}e^{i\psi}\cos\alpha\sin\alpha ,\nonumber\\
\rho_{44}(t) &=& e^{-2\kappa t}\sin^{2}\alpha , \label{e23}
\end{eqnarray}
from which we have for the concurrence
\begin{eqnarray}\label{28}
C(t) &=& {\rm max}\{0,\tilde{C}_{2}(t)\} \nonumber\\
&=& {\rm max}\!\left\{0,\!\left[|\sin2\alpha|\!-\!2(1\!-\!e^{-\kappa t})\!\sin^{2}\!\alpha\right]\!e^{-\kappa t}\right\} .
\end{eqnarray}
It is not difficult to see that $\tilde{C}_{2}(t)$ may reach a negative value at a finite time. This means that the effect of entanglement sudden death (ESD) can occur. The time evolution of $\tilde{C}_{2}(t)$ depends crucially on $\alpha$ and $\tilde{C}_{2}(t)$ is always positive when $\alpha<\pi/4$. For $\alpha>\pi/4$ the quantity~$\tilde{C}_{2}(t)$ can reach negative values at a finite $t$.

Notice from Eq.~(\ref{17}) that ESD is related to the population distribution between the states $\ket 1$ and $\ket 4$. Using Eq.~(\ref{e23}) we find
\begin{equation}
\rho_{11}(t) -\rho_{44}(t) = 1- 2e^{-\kappa t}\sin^{2}\alpha ,
\end{equation}
which shows that $\rho_{11}>\rho_{44}$ when $\alpha<\pi/4$ and $\rho_{11}<\rho_{44}$ for $\alpha>\pi/4$. Thus, no ESD is seen when the state $\ket 1$ is more populated than $\ket 4$.

Consider now the evolution of the double excitation $(n=2)$ states. In this case, the logarithmic negativity is employed to evaluate entanglement and one can show that for the initial NOON state~(\ref{e6}), the only negative eigenvalue of the partial transpose matrix $\rho^{T_{B}}$ is
\begin{equation}
    2\mu_{1} =\rho_{11}(t)-\sqrt{\rho_{11}^2(t)+|\rho_{37}(t)|^2} ,\label{30a}
\end{equation}
where
\begin{eqnarray}
        \rho_{11}(t) &=& 1-2e^{-\kappa t}+e^{-2\kappa t} , \nonumber\\
        \rho_{37}(t) &=& \rho_{73}^{\ast}(t) = e^{-2\kappa t}e^{i\psi} \cos\alpha\sin\alpha .
\end{eqnarray}
Clearly the eigenvalue (\ref{30a}) is always negative and the maximum negative value occurs for $\alpha=\pi/4$, i.e., for the initially maximally entangled NOON state. In other words, the modes remain entangled over the entire decay time and therefore the ESD never occurs.

For the initial EPR state, the negative eigenvalue of the transpose matrix can be determined from Eq.~(\ref{39}) as
\begin{equation}\label{30c}
    2\mu_{2} = \rho_{33}(t)\!+\!\rho_{77}(t)\!-\!\sqrt{[\rho_{33}(t)-\rho_{77}(t)]^2+4|\rho_{19}(t)|^2} ,
\end{equation}
where
\begin{eqnarray}\label{30d}
\rho_{33}(t) &=& \rho_{77}(t) = e^{-2\kappa t}\left(1-2e^{-\kappa t}+e^{-2\kappa t}\right)\sin^{2}\alpha ,\nonumber\\
\rho_{19}(t) &=& \rho_{91}^{\ast}(t) = e^{-2\kappa t}e^{i\psi}\cos\alpha\sin\alpha .
\end{eqnarray}
By substituting Eq.~(\ref{30d}) into Eq.~(\ref{30c}) one can easily obtain the explicit expression for $\mu_{2}$, which is
\begin{equation}\label{30u}
\mu_{2} = \left[\left(1-2e^{-\kappa t}+e^{-2\kappa t}\right)\sin^{2}\alpha -\frac{1}{2}|\sin2\alpha|\right]e^{-2\kappa t} .
\end{equation}
If $\alpha<\pi/4$, we see that $\mu_{2}$ is always negative and decreases exponentially with time. If $\alpha>\pi/4$ the eigenvalue may be positive at a finite time $t$.

We may conclude that both single and double excitation EPR states decaying under the influence of the ordinary vacuum field exhibit the ESD phenomenon whereas the NOON states decay exponentially and no ESD effect occurs.

\subsection{Dynamics in separate squeezed reservoirs}

We now turn to the problem of the evolution of an initial entanglement between the cavity modes each interacting with own separate reservoirs. Our interest will be centred principally on the manner the NOON and EPR states evolve in single-mode squeezed reservoirs. As is well known, a single-mode squeezed field contains strong two-photon correlations which are transferred to the cavity modes. An interesting question then arises, how these correlations could affect the evolution of the cavity modes already prepared in an entangled (correlated) state.
\begin{figure}[ht]
\center{{\bf (a)}\hskip5cm{\bf (b)}}
\resizebox{0.5\textwidth}{!}{%
\includegraphics{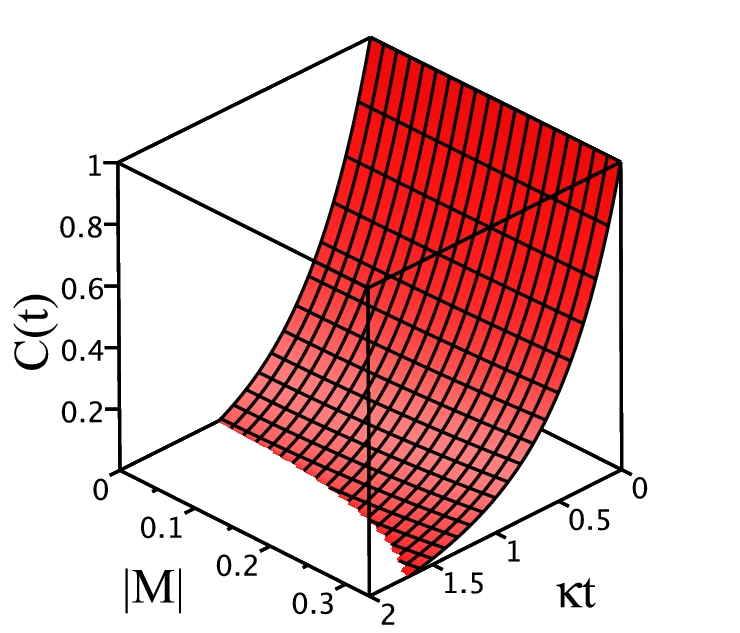}~\includegraphics{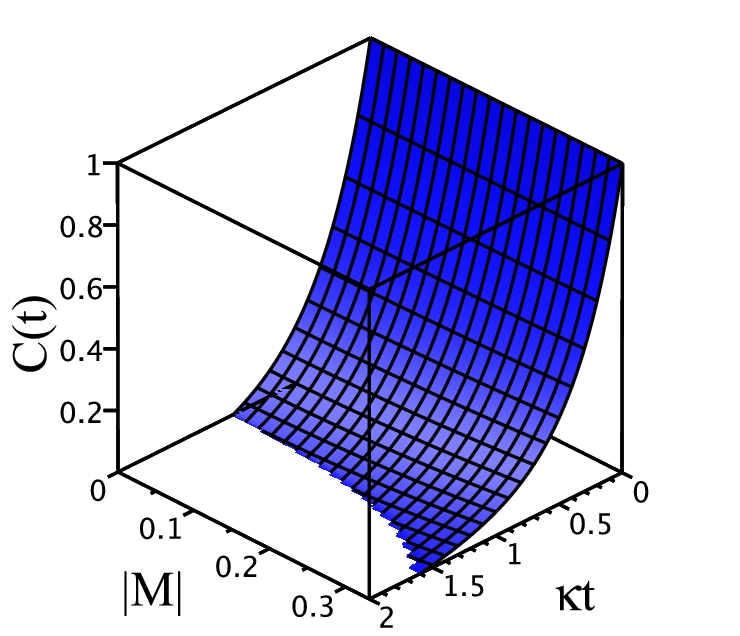}}
\caption{(Color online) Concurrence as a function of the dimensionless time $\kappa t$ and the correlation parameter $M$ for the initial single excitation (a) NOON and (b) EPR states of the cavity modes interacting with squeezed vacuum reservoirs with the mean photon number $N=0.1$ and relative phase $\theta=0$.}
\label{fig2}
\end{figure}

Figure~\ref{fig2} shows the time evolution of the concurrence of the cavity modes decaying into squeezed reservoirs and initially prepared in two different single excitation states, the NOON state (Fig.~\ref{fig2}a) and the EPR state (Fig.~\ref{fig2}b). We see that in both cases the sudden death time of the entanglement is almost unchanged for classically squeezed fields with $M<N$, which means that even if there are correlations in the reservoirs, but not quantum correlations, then the decay times of the initial entanglements remain almost the same as in the case of a thermal (uncorrelated, $M=0$) reservoirs. A modification of the decay time is observed for the correlation factor $M>N$, which corresponds to quantum squeezing. We may conclude that when the concurrences decay in quantum squeezed reservoirs one can expect significant changes in the decay time which can be regarded as a definite manifestation of the quantum nature of the squeezed reservoir.
\begin{figure}[h]
\center{{\bf (a)}\hskip4cm{\bf (b)}}
\resizebox{0.5\textwidth}{!}{%
\includegraphics{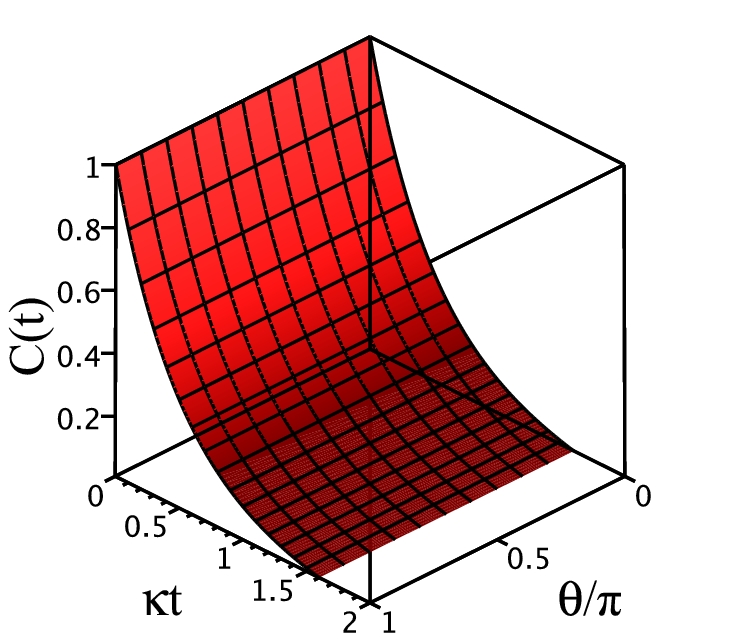}~\includegraphics{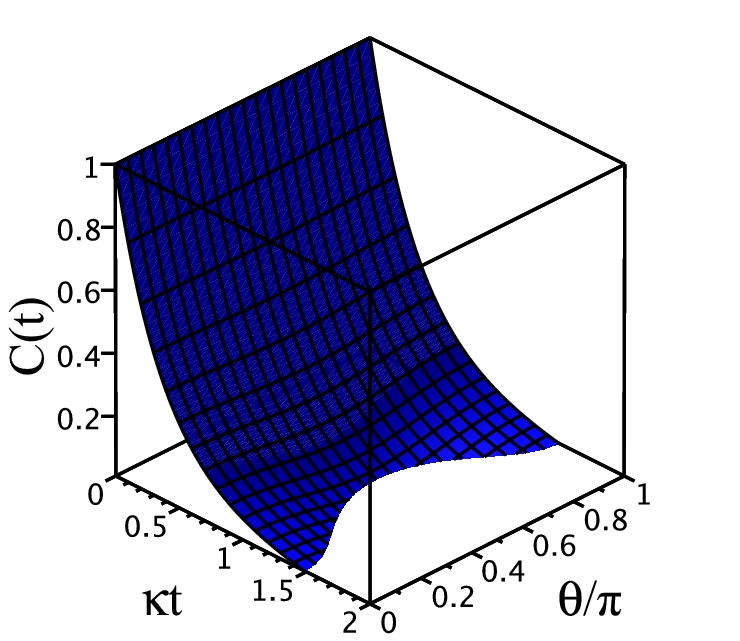}}
 \caption{(Color online) Concurrence as a function of the dimensionless time $\kappa t$ and the squeezing phase $\theta$ for the initial single excitation (a) NOON and (b) EPR states of the cavity modes interacting with squeezed vacuum reservoirs with the mean photon number $N=0.1$, the correlation parameter~$|M|=\sqrt{N(N+1)}$ and $\psi=0$.}
\label{fig3}
\end{figure}

One can notice from Fig.~\ref{fig2} that the time evolution of the concurrences of the NOON and EPR states in the squeezed reservoirs is almost identical. This is clear contrast to the case of the ordinary vacuum reservoirs where we have seen the concurrence for the two initial states decays in completely different manners. However, there are differences in the evolution in the squeezed reservoirs if one look at the phase properties of the concurrences. An example is shown in Fig.~\ref{fig3}. A comparison of Fig.~\ref{fig3}(a) and $3$(b) immediately shows the difference that for the initial NOON state, the evolution of the concurrence is completely independent of the phase while for the initial EPR state it exhibits a variation with the phase.

To examine the phase properties of the concurrence we look at the time evolution of the coherences $\rho_{23}$ and $\rho_{14}$. In the case of the initial NOON state, the coherence responsible for the entanglement is $\rho_{23}$. It is not difficult to find from Eq.~(\ref{e2}) that in the case of the initial state, Eq.~(\ref{e5}), the time evolution of the coherence is given by
\begin{eqnarray}
\rho_{23}(t) = \frac{1}{2} e^{i\psi}\sin(2\alpha)\cosh(2|M|\kappa t)e^{-(4N+1)\kappa t} .\label{32}
\end{eqnarray}
This equation shows explicitly that the coherence is independent of the squeezing phase $\theta$. Although the coherence depends on the phase~$\psi$, the absolute value of $\rho_{23}$ is independent of the phase. No phase dependence can thus be seen in the concurrence. The reason for this feature can be understood by noting that the NOON state is formed by a superposition of the states corresponding to a single excitation of either of the two modes $(\ket{0_{A},1_{B}}$ or $\ket{1_{A},0_{B}})$. As such, this cannot establish a fixed phase between the modes.

When the modes are initially prepared in the EPR state, Eq.~(\ref{e6}), the coherence responsible for entanglement is $\rho_{14}$, and then one can easy find that the time evolution of the coherence is given by
\begin{eqnarray}
\rho_{14}(t) &=& \frac{1}{2}\sin(2\alpha)e^{-(4N+1)\kappa t}e^{i\psi}\left[\cosh^{2}(|M|\kappa t)\right. \nonumber\\
&+&\left. e^{-2i(\theta+\psi)}\sinh^{2}(|M|\kappa t)\right] .\label{e32}
\end{eqnarray}
It is apparent that the coherence is composed of two terms among which only the second term depends on the relative phase of the squeezed fields and the initial state of the system. Hence, the absolute value too will depend on the squeezing phase $\theta$ and, therefore, the concurrence will vary with the phase. The presence of the phase dependence is linked to the fact that the EPR state is a superposition of the purely excited or de-excited states $(\ket{1_{A},1_{B}}$ or $\ket{0_{A},0_{B}})$, which corresponds to a simultaneous excitation or de-excitation of the two modes. The excitation and de-excitation processes can be completely random (incoherent) or they can occur simultaneously (coherently) with the fixed phase $\theta$. Alternatively, we may attribute the phase dependence to the fact that two empty $(\ket{0_{A},0_{B}})$ or excited $(\ket{1_{A},1_{B}})$ modes cannot be distinguished by a detection. Clearly, the incoherent and coherent processes are reflected in the time evolution of the coherence~$\rho_{14}$. The first term in the square bracket of Eq.~(\ref{e32}) represents the contribution to the coherence of the incoherent process whereas the second term, which depends on the squeezing phase, represents the contribution of the coherent process. Thus, a conclusion is worth making. Although the squeezed reservoir is characterised by the presence of phase dependent correlations, the case in which the concurrence depends on the phase occurs only for the initial EPR state. This fact suggests that by monitoring the evolution of an initial entangled state one could determine the state of unknown squeezed field.
\begin{figure}[hb]
\center{{\bf (a)}\hskip3.5cm{\bf (b)}}
\center{\includegraphics[height=0.45\columnwidth]{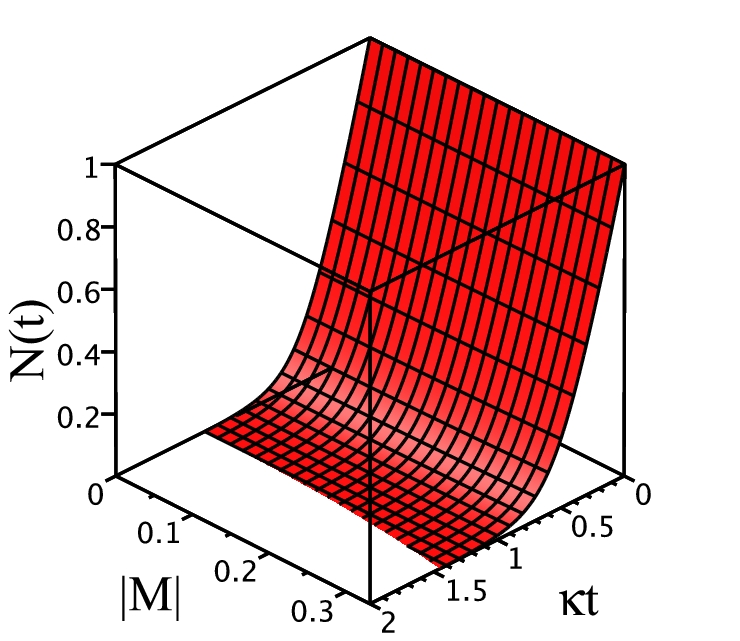}~\includegraphics[height=0.45\columnwidth]{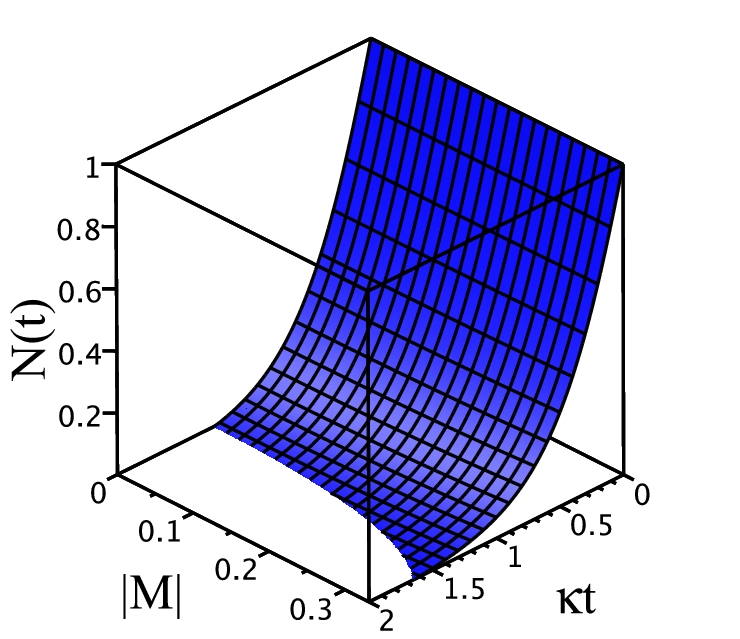}}
\caption{(Color online) test Variation of the logarithmic negativity with the dimensionless time $\kappa t$ and the correlation parameter~$|M|$ for the modes interacting with separate reservoirs with $N=0.1$, $\psi=\theta=0$ and the initial single excitation $(n=1)$ (a) NOON and (b) EPR states.}
\label{fig4}
\end{figure}

We have limited our considerations of the dynamics of the entangled states in separate squeezed reservoirs to the case of single excitation states only. Similar conclusions can be reached for cases involving entanglement of the doubly excited NOON and EPR states. Figure~\ref{fig4} shows the dependence of the decay time of the initial entanglement on the correlation parameter $|M|$. It is seen that similar to the case of single excitation states, shown in Fig.~\ref{fig2}, the decay time of the negativities is not sensitive to $|M|$ until $|M|>N$. Thus, we are clearly able to distinguish nonclassical effects from classical ones. One notable difference between the dynamics of the singly and doubly excited states is that the decay time of the entanglement of the doubly excited NOON state, Fig.~\ref{fig4}(a), is less sensitive to $|M|$ than the corresponding singly excited state, shown in Fig.~\ref{fig2}(a).

\subsection{Dynamics in a common squeezed reservoir}

We now turn to the case when the cavity modes interact with a common squeezed reservoir and study the dynamics of entanglement of the initial single and double excitation NOON and EPR states. As before for the separate reservoirs, we search for signatures of quantum correlations present in the squeezed reservoir. Also, we discuss the role of the correlations present in the squeezed reservoir in the evolution of the initial correlations between the modes and note the importance of the relative phase between the squeezed field and the initial entangled state of the system.

\subsubsection{Singly excitation states}

\begin{figure}[h]
\center{{\bf (a)}\hskip4cm{\bf (b)}}
\resizebox{0.5\textwidth}{!}{%
\includegraphics{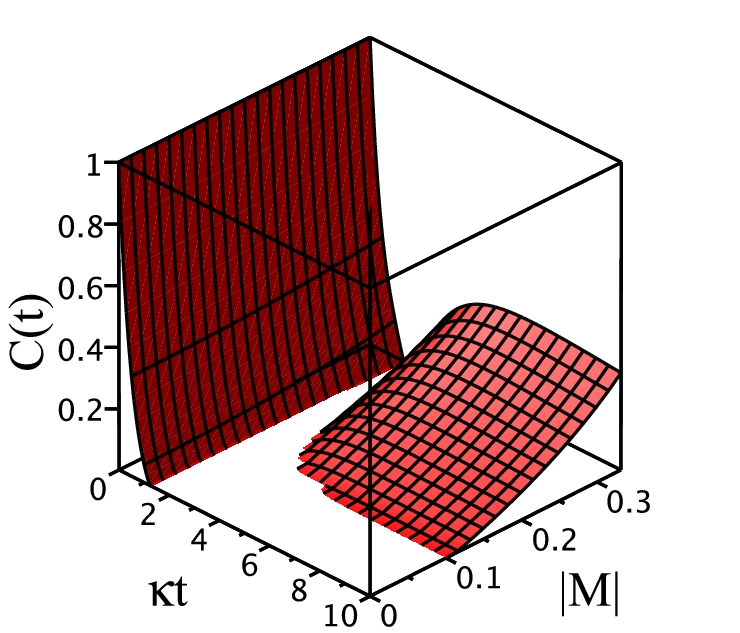}~\includegraphics{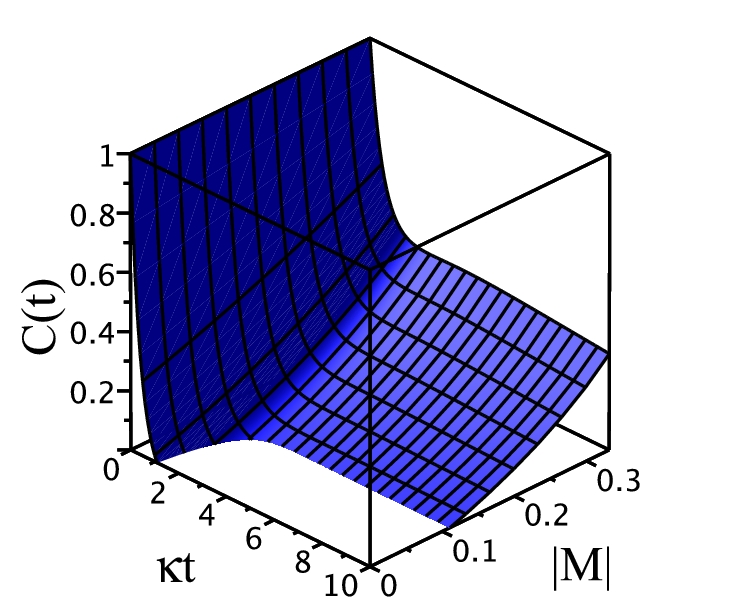}}
\caption{(Color online) Concurrence as a function of the dimensionless time $\kappa t$ and the correlation parameter $M$ for the initial single excitation (a) NOON and (b) EPR states of the cavity modes interacting with a common squeezed reservoir with the mean photon number $N=0.1$ and relative phase $\theta=0$.}
\label{fig5}
\end{figure}

Consider first the evolution of the single excitation entangled states. Figure~\ref{fig5} shows the concurrence as a function of time and the correlation parameter $|M|$ for the initial single excitation NOON (Fig.~\ref{fig5}(a)) and EPR (Fig.~\ref{fig5}(b)) states. Similar to the previous case of separate reservoirs, the initial entanglement decays quite rapidly in time and disappears over a short time. However, in contrast to the case of separate reservoirs, as time progresses an entanglement reappears again and then remains to the steady state limit.

We see from the figure that in both cases the evolution of the entanglement can be divided into two regions of different behaviour. The first region corresponds to that of short times in which the initial entanglement decays rapidly in time. The second region refers to longer times at which an revival of entanglement is observed or a significant slow down of the decay process leading to a non-zero entanglement for all times.

However, the most prominent feature of these results is that the revival of entanglement occurs only if the reservoir is quantum squeezed $(|M|>N)$. This applies to both initial states. Thus, the revival of entanglement is associated with quantum aspects of the squeezed reservoir. This is of course a reflection of the fact that entanglement is a nonclassical feature associated with quantum correlations. Therefore, the revival of entanglement clearly corresponds to a transfer of quantum correlations from the squeezed reservoir to the modes of the two cavities.

Notice that the decay process of entanglement of the initial NOON state undergoes the ESD effect that it disappears at a finite time and then reappears again to remain nonzero for all times. However, the behaviour of the concurrence of the initial EPR state is quite different. We see that the decay of the initial entanglement is much slower than an exponential decay and the entanglement is preserved even after a long time.
\begin{figure}[h]
\center{\includegraphics[width=0.8\columnwidth]{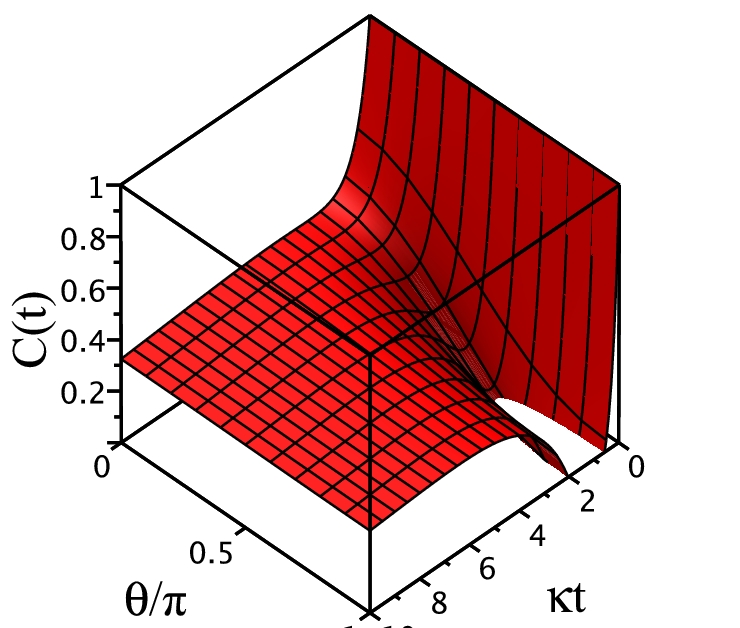}}
\caption{(Color online) Concurrence as a function of the dimensionless time $\kappa t$ and the squeezing phase $\theta$ for the initial single excitation EPR state of the cavity modes $(\alpha=\pi/4, \psi=0)$ interacting with a common squeezed reservoir with the mean photon number $N=0.1$ and $|M|=\sqrt{N(N+1)}$.}
\label{fig6}
\end{figure}

In order to explain that feature, we plot in Fig.~\ref{fig6} the variation of the concurrence with time and the phase $\theta$ of the squeezed field. We have made a choice of the phase of the initial state $\psi=0$, so that the relative phase between the initial state and the squeezed field is controlled through the phase $\theta$. We see that the evolution of the concurrence near $\kappa t\approx 1$ changes qualitatively when $\theta$ is varied. For $\theta <\pi/2$ entanglement is seen to occurs over the entire evolution time. The effect of increasing $\theta$ to $\theta=\pi$ leads to the ESD and then to entanglement revival. Note that the phase of the initial EPR state was~$\psi=0$. We may conclude that the entanglement persists in the system for all times when the initial EPR state and the state of the squeezed reservoir are in phase $(\theta-\psi=0)$ but it undergoes the ESD effect when the states are in the opposite phase $(\theta-\psi=\pi)$. This phase difference is just the origin of the ESD feature seen in Fig.~\ref{fig6}. Therefore, the relation between phase of the initial state and the phase of the squeezed field plays an important role in the evolution of entanglement. The strong variation of the concurrence with $\theta$ at $\kappa t\approx 1$ can be interpreted as resulting from a constructive $(\theta-\psi =0)$ and destructive $(\theta-\psi =\pi)$ interference between the initial entangled state and that of the squeezed reservoir.
\begin{figure}[h]
\center{{\bf (a)}\hskip4cm{\bf (b)}}
\resizebox{0.45\textwidth}{!}{%
\includegraphics{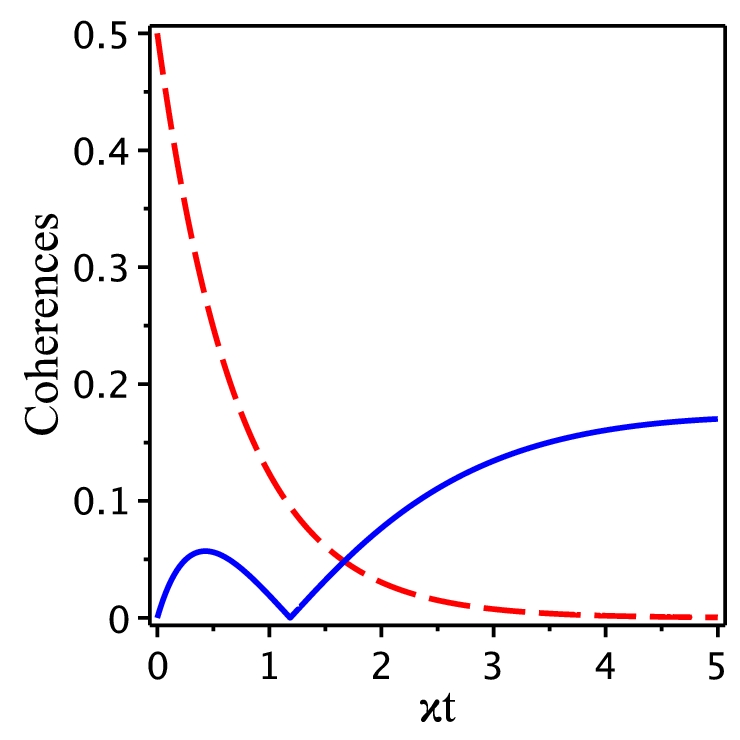}~\includegraphics{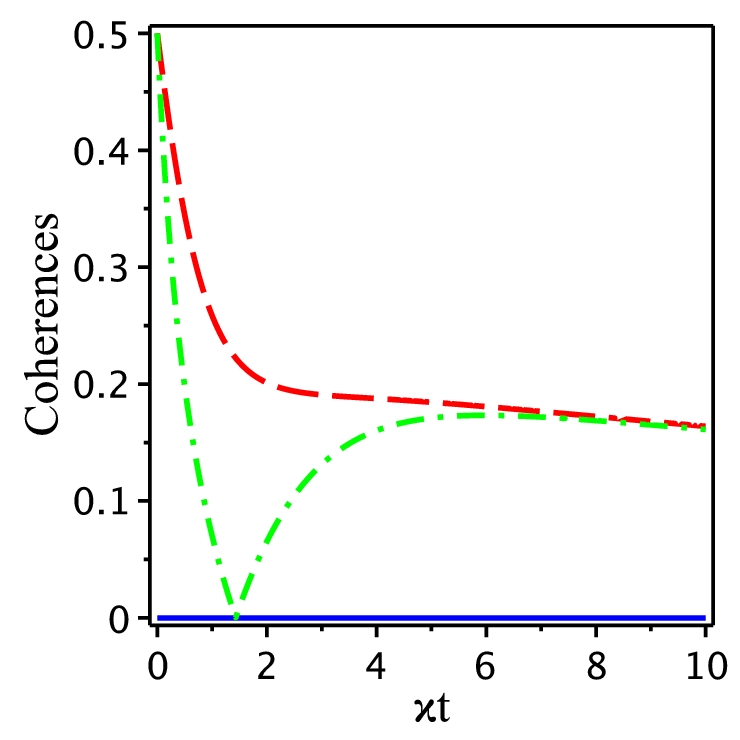}}
\caption{(Color online) Time evolution of the coherences $\rho_{14}$ (solid blue line) and $\rho_{23}$ (dashed red line) for $\theta=0$ and different initial (a) NOON and (b) EPR states of the cavity modes interacting with a common squeezed reservoir. The phase of the initial state is $\psi=0$, the mean photon number $N=0.1$ and $|M|=\sqrt{N(N+1)}$. Also shown in (b) by dashed-dotted green line is the coherence $\rho_{14}$ for $\theta =\pi$.}
\label{fig7}
\end{figure}

It is interesting to distinguish the difference between the state of the revival entanglement and that of the initial entanglement. In the case of single excitation there are two coherences that could create entangled states, $\rho_{23}$ which creates the NOON state and $\rho_{14}$ which creates the EPR state. Figure~\ref{fig7} shows the time evolution of the coherences $\rho_{23}$ and $\rho_{14}$ for different initial states with $\psi=0$, (a) NOON state and (b) EPR state. In the case (b), it also shows the variation of the coherence $\rho_{14}$ with the phase $\theta$. Two things are immediately apparent in Fig.~\ref{fig7}. The initial coherence is a rather a short lived affair, but it lives longer than the initial entanglement. Also, it is quite apparent that the coherence $\rho_{14}$ dominates over longer times and remains nonzero until the steady state. Thus, the decay of the initial entanglement is solely due to the decay of the initial coherences and the appearance of the long time entanglement is entirely due to the coherence $\rho_{14}$. In other words, the long time entanglement is associated with an EPR state. Comparing the evolution of the coherences (Fig.~\ref{fig7}) with the evolution of the concurrences (Fig.~\ref{fig5}) one can notice that the coherence lives longer than the initial entanglement and recurrence of entanglement occurs after the initial coherence vanishes.

\subsubsection{Doubly excitation states}

We turn now to a discussion of the time evolution and transfer of entanglement of the double excitation states, where up to two excitations can be present in each mode. The logarithmic negativity, which is adopted to quantify entanglement, is readily calculated from Eq.~(\ref{19}).
\begin{figure}[h]
\center{{\bf (a)}\hskip3.5cm{\bf (b)}}
\center{\includegraphics[height=0.45\columnwidth]{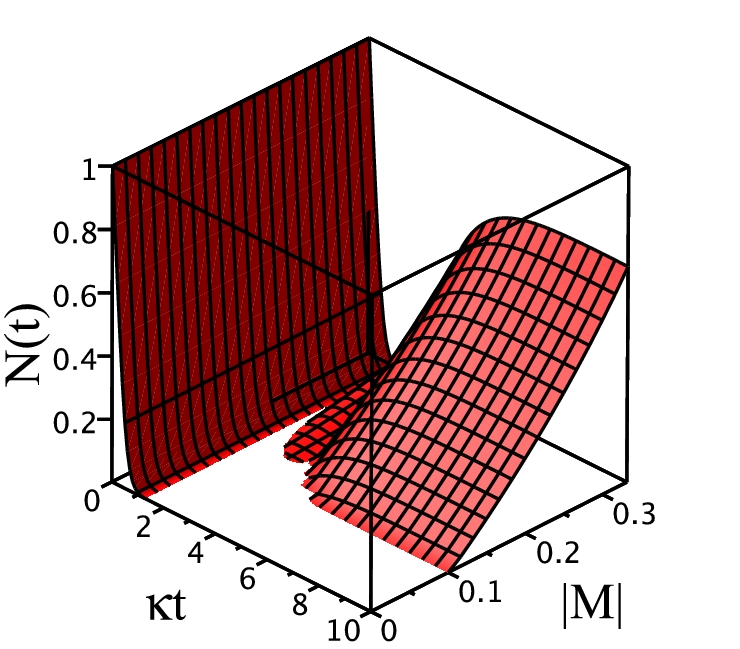}~\includegraphics[height=0.45\columnwidth]{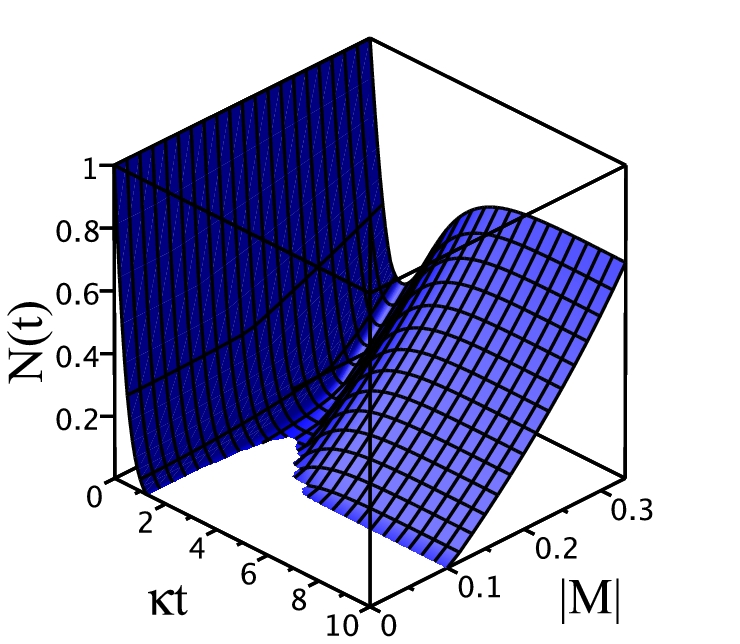}}
\caption{(Color online) Variation of the logarithmic negativity with the dimensionless time $\kappa t$ and the correlation parameter~$|M|$ for the modes interacting with a common reservoir with $N=0.1$, $\psi=\theta=0$ and the initial double excitation (a) NOON and (b) EPR states.}
\label{fig8}
\end{figure}
\begin{figure}[h]
\center{\includegraphics[height=0.3\textwidth]{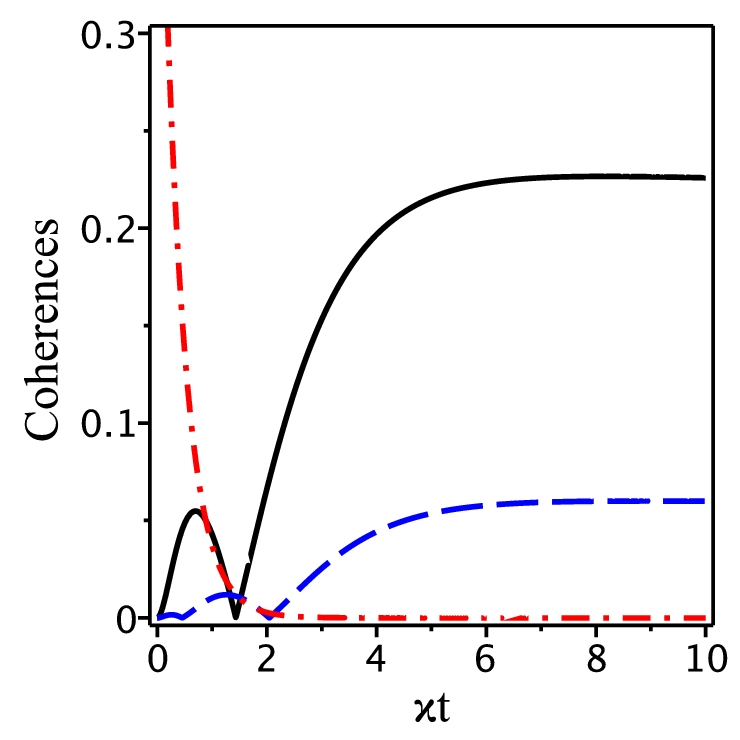}}
 \caption{(Color online) Time evolution of the coherences $\rho_{15}$ (solid black line), $\rho_{19}$ (dashed blue line) and $\rho_{37}$ (dashed-dotted red line) for the initial NOON state with $\alpha=\pi/4$ and $\psi=0$. The parameters of the squeezed field are $N=0.1$, $|M|=\sqrt{N(N+1)}$ and $\theta=0$.}
\label{fig9}
\end{figure}
Figure~\ref{fig8} shows the variation of the logarithmic negativity with time and the correlation parameter $|M|$ for the two different initial states of the system. The behaviour of the logarithmic negativity is seen to be qualitatively similar to the concurrence shown in Fig.~\ref{fig5} for the single excited states. Again, there is no entanglement revival for $|M|\leq N$. As $|M|$ increases above $N$, the revival of entanglement occurs. One can notice that the revival of entanglement for the initial NOON state occurs without a time delay, in contrast to the previous case of the singly excited states, Fig.~\ref{fig5}(a). The reason is that in the present case, where up to two photons can be absorbed by each of the mode, both singly and doubly excited EPR states can be generated during the evolution. This is shown in Fig.~\ref{fig9} which illustrates the time evolution of the coherence $\rho_{37}$ responsible for entanglement of the initial state, and coherences $\rho_{15}$ and $\rho_{19}$ responsible, respectively, for the generation of the singly and doubly excited EPR states. The coherences $\rho_{15}$ and $\rho_{19}$ are associated with quantum correlations transferred to the modes from the squeezed reservoir. Clearly, the revival of entanglement without a delay, seen in Fig.~\ref{fig8}(a), follows the combined evolution of the coherences $\rho_{15}$ and~$\rho_{19}$.

\section{Conclusion}\label{sec5}

We have studied the transient properties of entanglement in a bipartite system composed of two independent single-mode cavities. We have examined the cases when each cavity is coupled to a squeezed reservoir and when both cavities are coupled to a common squeezed reservoir. We have illustrated our considerations by examining time evolution of entanglement of initial single and double excitation NOON and EPR states. Characterising the squeezed reservoir by the correlation parameter $|M|$ and the phase~$\theta$, we have identified feature that are unique to nonclassical (quantum) nature of the squeezed reservoir that they take place only when $|M|>N$, i.e. when the reservoir correlations exceed the number of photons $N$ in the squeezed field. A classically squeezed field is that with $|M|\leq N$. We have found that in the case of the interaction with separate reservoirs the initial entanglement of the cavity modes disappears at a finite time which can be influenced only if the modes interact with quantum squeezed fields. For the so-call classically squeezed field, the decay time of the initial entanglement remains almost the same as in the case of a thermal field. We have also shown that only EPR states exhibit a dependence on the squeezing phase, and the entanglement also occurs in a less restricted range of the evolution time. When the cavity modes interact with a common reservoir, a revival of entanglement may occur. It is found that this feature is also unique to quantum squeezing. There is no entanglement revival when the modes interact with a classically correlated squeezed field. The revival of the entanglement has be interpreted as a transfer of quantum correlations from the squeezed reservoir to the cavity modes. The effect of initial correlations on the entanglement transfer from the squeezed reservoir to the modes has also been analysed. It has been found that initially entangled modes with correlations different from those present in the squeezed reservoir lead to a delay of the transfer of the correlations from the reservoir to the modes. Moreover, the initial coherence can live longer than the initial entanglement and the revival of entanglement occurs after the initial coherence vanishes.

\acknowledgments

The authors KA and SB acknowledge grant support from the Dean for
Scientific Research of Taibah University (Grant number 3079/34).

%\begin{acknowledgement}
%\textbf{Acknowledgement}\\
%The authors KA and SB acknowledge grant support from the Dean for Scientific Research of Taibah University (Grant number 3079/34).
%\end{acknowledgement}
% end of file template.tex

\end{document}